# A Tolman Surface Brightness Test for Universal Expansion, and the Evolution of Elliptical Galaxies in Distant Clusters[1,2]

Michael A. Pahre[3], S. G. Djorgovski[3,4], and R. R. de Carvalho[3,5]

## ABSTRACT

We use the intercept of the elliptical galaxy radius–surface brightness (SB) relation at a fixed metric radius as the standard condition for the Tolman SB test of the universal expansion. We use surface photometry in the optical and near-IR of elliptical galaxies in Abell 2390 ($z = 0.23$) and Abell 851 ($z = 0.41$), and compare them to the Coma cluster at $z \approx 0$. The photometric data for each cluster are well-described by the Kormendy relation $r_e \propto \Sigma_e^A$, where $A = -0.9$ in the optical and $A = -1.0$ in the near-IR. The scatter about this near-IR relation is only 0.076 in $\log r_e$ at the highest redshift, which is much smaller than at low redshifts, suggesting a remarkable homogeneity of the cluster elliptical population at $z = 0.41$. We use the intercept of these fixed-slope correlations at $R_e = 1$ kpc (assuming $H_0 = 75$ km s$^{-1}$ Mpc$^{-1}$, $\Omega_0 = 0.2$, and $\Lambda_0 = 0$, where the results are only weakly dependent on the cosmology) to construct the Tolman SB test for these three clusters. The data are fully consistent with universal expansion if we assume simple models of passive evolution for elliptical galaxies, but are inconsistent with a non-expanding geometry (the tired light cosmology) at the $5\sigma$ confidence level at $z = 0.41$. These results suggest luminosity evolution in the restframe $K$-band of $0.36 \pm 0.14$ mag from $z = 0.41$ to the present, and are consistent with the ellipticals having formed at high redshift. The SB intercept in elliptical galaxy correlations is thus a powerful tool for investigating models of their evolution for significant lookback times.

*Subject headings:* cosmology: observations — galaxies: evolution — galaxies: elliptical

## 1. Introduction

Tolman (1930; 1934) and Hubble & Tolman (1935) proposed the dimming of surface brightness (SB) with redshift as a test of the universal expansion. In an expanding universe, bolometric SB will decrease with redshift as $(1+z)^{-4}$, while in a non-expanding geometry, such as a tired light cosmological model where some property other than expansion gives rise to redshift, SB will decrease as $(1+z)^{-1}$, independent of all other cosmological parameters.

The key obstacle to performing this test is the definition of a standardized unit of SB which can be observed at a range of redshifts. Sandage & Perelmuter (1990a,b, 1991) have made the most thorough study to date, but mostly relied on poor quality photometry from the literature, and used the SB–luminosity correlation which has a very large intrinsic scatter. A better approach is to use the SB intercepts of the sharp Fundamental Plane (FP) correlations for elliptical galaxies, since these show the least scatter (Djorgovski & Davis 1987; Dressler *et al.* 1987; see Djorgovski 1992 for a review and references). For a discussion of this methodology as applied to the Tolman SB test, see Kjærgaard, Jørgensen, & Moles (1993); a related study was presented by Franx (1993).

We have adopted this methodology in principle, but in the absence of velocity dispersion measurements for ellipticals at larger redshifts, we have chosen to utilize a projection of the FP, the Kormendy relation (Kormendy 1977) between the effective radius $r_e$ and the mean SB $\langle\mu\rangle_e$ enclosed by that radius. We use the intercept of this relation at $R_e = 1$ kpc as the standard condition. The results of any Tolman SB test where galaxies must be corrected to a standard condition will involve some dependence on the assumed cosmology, but as will be described below for the redshifts of interest here, the effect of cosmology is quite small compared to the predicted difference between the expansion and tired light models. At larger redshifts, galaxy evolution may become important, as will the assumed cosmological parameters, so the Tolman SB test is best performed at low to moderate redshifts. A preliminary report of the data found in this *Letter* can be found in Djorgovski, Pahre, & de Carvalho (1995).

## 2. The Data

We observed 15 galaxies in the Coma cluster ($z = 0.024$) in the $K$-band using a new near-infrared camera (Murphy *et al.* 1995) on the Palomar Observatory 1.52 m telescope; the reductions and analysis are given in Pahre, Djorgovski, & de Carvalho (1995; hereafter PDdC95). We have also measured $r_{\eta=2}$ and $\langle\mu_K\rangle_{\eta=2}$, where $\eta = 2$ represents $-2.5\log$ of the ratio of the isophotal SB to the mean enclosed SB (each in linear units), for the Petrosian (1976) $\eta$ function. We have added observations of elliptical galaxies in the Coma cluster in the $B$- and $R_C$-bands (from Jørgensen, Franx, & Kjærgaard 1995, where we have utilized their relation to convert the observed $r$-band to $R_C$-band). We have removed the $k$-correction and SB dimming correction that were applied to the measurements of $\langle\mu\rangle_e$ for both the optical and near-infrared observations.

Observations in the $K$-band of a $45'' \times 60''$ field with a $0.15''$ pixel size near the core of cluster Abell 2390 ($z = 0.23$) were made with the near-infrared camera (Matthews & Soifer 1994) on the W. M. Keck 10 m telescope on 1994 October 16. The seeing was estimated at $0.45''$ from focus and standard stars taken before and after the cluster, and from one possible star in the cluster field-of-view (FOV). The integration time was 1200 s in two overlapping pointings. Observations of stars from Casali & Hawarden (1992) were used for calibration. We have used galaxies in the field that appear to follow a de Vaucouleurs $a^{1/4}$ profile (de Vaucouleurs 1953). Since most the effective radii were $< 1''$, we have chosen to utilize the Petrosian (1976) $\eta$ function with the value $\eta = 2$, which corresponds to the radius enclosing three-quarters of the light for a galaxy obeying a de Vaucouleurs profile. We have corrected $\langle\mu\rangle_{\eta=2}$ for seeing effects on the FP intercept using the simulations of Kjærgaard et al. (1993). All of the corrections are $< 0.1$ mag, and we estimate that the uncertainty of these seeing corrections is $< 0.02$ mag.

We have used public early-release HST/WFPC-2 observations of Abell 851 from 1994 January, representing approximately 20,000 s taken in the filter F702W. These data were reduced in the usual way using the STSDAS package of IRAF. We have made a charge transfer efficiency (CTE) correction for pre–1994 April data as given in Holtzman et al. (1995), and have made the appropriate distortion correction to measured scales and total magnitudes as described there. We have also observed the cluster in the Cousins $B$-, $R_C$-, and $I_C$-bands using a thinned, high quantum efficiency $2048 \times 2048$ pixel CCD on the Palomar 1.52 m telescope. We used standard stars from Landolt (1992) to establish the zero-point and color terms of the ground-based data, and then chose four of the largest elliptical galaxies to calibrate the HST F702W data onto the $R_C$-band system by comparing curves of growth at large radii. We conservatively estimate a 0.062 mag total error in our derived zero-point for the HST image for the color range provided by the four galaxies ($2.5 < B - R_C < 3.0$).

We have also imaged Abell 851 in the $K_s$-band with the Palomar 1.52 m telescope on UT 1995 February 11 with the same setup, reduction, and calibration as for the Coma galaxies. We integrated 5100 s using 60 s individual exposures, and moved the telescope $10-15''$ between each sequence of $5 \times 60$ s.

In Abell 851 we have used the $BR_CI_C$ data to select probable cluster elliptical galaxies by identifying the early-type galaxy locus in the color-magnitude, color-color, and concentration index–SB (Abraham et al. 1994; Fukugita et al. 1995a) diagrams; the method will be described in detail elsewhere (Pahre 1995). We have added to this list the spectroscopically-confirmed cluster elliptical galaxies from Dressler & Gunn (1992). We measured the effective radii $r_e$ and the mean surface brightnesses $\langle\mu\rangle_e$ enclosed by those radii for each galaxy from the HST image. We have excluded any galaxies with poor fits to a de Vaucouleurs $a^{1/4}$ profiles, and we have also excluded any galaxies which are disk-dominated as determined visually from the HST image. For the 15 remaining galaxies, we then converted the observed $\langle\mu_{R_C}\rangle_e$ measurements from the HST image into $\langle\mu_B\rangle_e$ and $\langle\mu_K\rangle_e$ using the $(B - R_C)$ and $(R_C - K)$ colors measured in a $4.5''$ diameter circular aperture from the Palomar 1.52 m data.

We have corrected all data for foreground galactic extinction based on the maps of Burstein & Heiles (1982).

## 3. The Tolman SB Test

As described in the previous section, we have constructed a sample of three rich clusters for our Tolman SB test: Coma, Abell 2390, and Abell 851. We have measurements of $r_e$ and $\langle\mu\rangle_e$ for the first and last clusters in the observed $B$-, $R_C$-, and $K$-bands. For Coma and Abell 2390, we have measurements of $r_{\eta=2}$ and $\langle\mu_K\rangle_{\eta=2}$; as described below, we will use these measurements to convert $r_{\eta=2}$ and $\langle\mu_K\rangle_{\eta=2}$ into $r_e$ and $\langle\mu_K\rangle_e$ for Abell 2390.

The Coma cluster $K$-band surface photometry for 15 galaxies from PDdC95 is best-fit by the Kormendy SB–radius relation to be $r_e \propto I_e^{-0.83\pm0.15}$, where the uncertainty represents solely the formal fitting error, with a quartile-estimated scatter about this relation of 0.17 in $\log r_e$. The Abell 851 data, however, are best-fit by a value of the exponent of $-0.93 \pm 0.07$ with a very small scatter of 0.076 in $\log r_e$; this scatter is increased by 50% if we used the exponent of $-0.83$ instead. The tightness of this relation for Abell 851 is readily apparent in Figure 1, which can be contrasted with Figure 1 of PDdC95. For the purposes of this *Letter*, we will assume the relation of $r_e \propto \Sigma_e^{-1.0}$ for the $K$-band (which has identical scatter to that with the exponent of $-0.93$), and will determine the effects on the SB test that this assumption entails.

All projected radii $r_e$ and $r_{\eta=2}$ have been converted into physical radii $R_e$ and $R_{\eta=2}$ by assuming the cosmology $H_0 = 75$ km s$^{-1}$ Mpc$^{-1}$, $\Omega_0 = 0.2$, and $\Lambda_0 = 0$. We have measured the median intercept for each cluster where $\log R = 0$, and estimated the rms using the quartiles of the residual distribution. For Abell 2390, we have measured the difference between the $\langle\mu\rangle_{\eta=2}$ intercept for this cluster and the Coma cluster, assumed that the difference will be identical for both the $R_e$–$\langle\mu\rangle_e$ and the $R_{\eta=2}$–$\langle\mu\rangle_{\eta=2}$ relations, and thus converted the $\langle\mu\rangle_{\eta=2}$ intercept into the $\langle\mu\rangle_e$ intercept. We note that this is a 0.21 mag arcsec$^{-1}$ correction. The results for all clusters are given in Table 1, where the errors reflect the contributions of the number of galaxies observed per cluster ($\sigma/\sqrt{N_{gal}}$), the median photometric errors in the ($R_C - K$) color (Abell 851 only), the CTE effect and distortion correction (Abell 851 only), the zero-point uncertainty of 0.03 mag, and the uncertainty in the relative offset between the $r_e$–$\langle\mu\rangle_e$ and $r_{\eta=2}$–$\langle\mu\rangle_{\eta=2}$ relations in the Coma cluster (the latter for Abell 2390 only).

We have plotted in Figure 2 the results for the $K$-band which comprise our Tolman SB test. For comparison, we have also plotted a number of models. We include an expanding universe where SB goes as $(1 + z)^{-4}$, with and without the $k$-correction from Buzzoni (1995); a tired light model where SB goes as $(1 + z)^{-1}$, with and without $k$-correction; and evolution models (Models 1 and 2, where they differ in values of the cosmological parameters) from Buzzoni (1995), for which we have added the universal expansion SB dimming signal. All models have been normalized to the Coma cluster data point. It is apparent from the figure that the SB dimming between the three clusters in the $K$-band is best represented by the evolution models with the expected universal expansion model, while the two cosmological models cannot be distinguished. The measured dimming in the $K$-band between the Coma ($z = 0.024$) and Abell 851 ($z = 0.407$) clusters is $0.73 \pm 0.14$ mag. We expect to find a differential SB dimming of 1.38 mag due to the expansion, and the differential $k$-correction of Buzzoni (1995) is $-0.29$ mag resulting in an observed discrepancy of $0.36 \pm 0.14$ mag out to a redshift of $z = 0.407$. The data thus exclude the no-evolution expansion model at the $2.6\,\sigma$ level, and the tired light no-evolution model at the $> 5\,\sigma$ level. The data would be consistent with universal expansion, however, if we require $0.36 \pm 0.14$ mag of luminosity evolution in the $K$-band out to $z = 0.407$. From Figure 2, we see that the result for Abell 2390 is fully consistent with that for Abell 851 above, albeit at lower significance, hence the assumptions we used to convert $\langle \mu \rangle_{\eta=2}$ to $\langle \mu \rangle_e$ have not compromised our final result for Abell 2390.

If we choose to use the cosmologies ($H_0$, $\Omega_0$, $\Lambda_0$) of (50, 1, 0) or (100, 0.01, 0) in converting from angle to physical scale, then the relative dimming between the Coma and Abell 851 clusters (in the sense of $\langle \mu \rangle_e^{\text{Coma}} - \langle \mu \rangle_e^{\text{A851}}$) changes by $-0.07$ and $-0.09$ mag arcsec$^{-2}$, respectively. If we choose to use $r_e \propto \Sigma_e^{-1.1}$ or $r_e \propto \Sigma_e^{-0.9}$, then the same relative dimming changes by $+0.08$ and $-0.11$ mag arcsec$^{-2}$, respectively, where the effect is primarily due to the tight $r_e$–$\langle \mu \rangle_e$ relation for Abell 851. If the Coma cluster possesses a $-200$ km s$^{-1}$ peculiar velocity, then the result would change by $+0.03$ mag arcsec$^{-2}$. The measured SB dimming is thus only weakly dependent upon the specific assumptions about slope, cosmology, or peculiar velocity.

When constructing the Tolman SB test in the optical, we have found that the observations of both the Coma and Abell 851 clusters are well-described by a relation of the form $R_e \propto \Sigma_e^{-0.9}$. The median intercepts at log $R_e = 1$ kpc for the $B$- and $R_C$-bands are given in Table 1 and plotted in Figure 3. We plot comparison models as in Figure 2, with the exception that the $k$-correction (i.e. no evolution) models for the $R_C$-band come from Fukugita, Shimasaku, & Ichikawa (1995b). We note that the differential $k$-correction between the $r$-band for the Buzzoni (1995) models and our observed $R_C$-band is quite small for the relevant redshifts, as is shown explicitly by Fukugita et al. (1995b). The $R_C$-band Tolman SB test in Figure 3 shows a similar result to the $K$-band, as the result is consistent with passive evolution of the elliptical galaxy population. The $B$-band Tolman SB test appears to depart from the passive evolution models; we note, however, that the required evolution of $0.34 \pm 0.20$ mag to $z = 0.407$ differs from the evolution model ($H_0 = 100$ km s$^{-1}$ Mpc$^{-1}$, $\Omega_0 = 0$) prediction of 0.64 mag by only $0.30 \pm 0.20$ mag, which is not of high significance. A comparison of the models in Figures 2 and 3 emphasizes the advantage of constructing the Tolman SB test in the $K$-band, as there is a larger separation between the no-evolution tired light model and the evolution with expansion model.

## 4. Discussion and Conclusions

The Tolman SB test constructed using the Kormendy relation at these redshifts is relatively insensitive to the choice of cosmology, the exact exponent of the relation, or small peculiar velocities for the nearby calibration cluster. Using our adopted slope of $-1.0$, we measure a $0.73 \pm 0.14$ mag arcsec$^{-2}$ dimming which excludes the tired light model at the $> 5\,\sigma$ level. Since the Tolman SB test is a direct consequence of the geometry of expanding spacetime, the $K$-band Tolman SB test constructed in this *Letter* is therefore the strongest case yet made on geometrical grounds for the exclusion of a non-expanding cosmological model. In the modification of the Tolman SB test we have used for this *Letter*, however, we have established the SB intercept of the elliptical galaxy $r_e$–$\langle\mu\rangle_e$ correlation as a standardized unit of SB at the expense of losing the strict independence of the test from differences between cosmological models.

Our observations in the $K$-band suggest that there has been luminosity evolution since $z \sim 0.4$ in the cluster elliptical galaxy population. This result was obtained for a limited dataset for the local calibration of the Coma cluster (only 15 galaxies), a single high-redshift cluster, and a higher-scatter projection of the FP. We are undertaking a large survey which will improve on all three of these limitations, thereby vastly increasing the significance of evolution in the cluster elliptical galaxy population within this redshift range. Numerical simulations suggest that hierarchical merging tends to move galaxies along, not perpendicular to, the FP (Capelato, de Carvalho, & Carlberg 1995), therefore we do not expect our results to be affected by dynamical evolution. The observed SB dimming is consistent with the predictions of Buzzoni's population models with evolution, but the accuracy of our results is clearly not sufficient to distinguish between different values of the cosmological parameters, including the formation redshift. The tightness of the $r_e$–$\langle\mu\rangle_e$ correlation for Abell 851, and the relatively small amount of luminosity evolution required at $z = 0.41$, nevertheless suggest that the formation epoch for ellipticals in rich clusters at intermediate redshifts was $z > 1$. Ellis (1995) has argued that observational constraints on color evolution for the cluster elliptical population out to $z = 0.55$ must be less than 0.07 mag in rest-frame $(U - V)$. Such small color evolution likewise suggests that the formation redshift for cluster ellipticals was $z > 3$, which is fully consistent with our results. We note that our results are also qualitatively consistent with those presented by Rigler & Lilly (1994) and Dickinson (1995).

The detection of luminosity evolution has been suggested in the past, based upon Hubble diagrams of radio galaxies (see Sandage 1988, and references therein). The advantages of the approach we have used in this *Letter* are that we avoid aperture corrections, do not rely on the radio galaxy population (which may be evolving more strongly), and can work at lower redshifts where evolutionary effects and differences between cosmological models are smaller. While there is no surprise in our results which exclude the non-expanding universal geometry at high confidence, the unprecendented accuracy with which we have been able to measure the SB dimming (and hence luminosity evolution) using a limited and heterogeneous dataset is highly significant. Future work probing the evolution of elliptical galaxies with redshift, and possible variations in the slopes of elliptical galaxy correlations, should provide a deeper understanding of the formation of elliptical galaxies in rich clusters.

Thanks are due to the builders of the near-infrared instrumentation at the Palomar 60" (D. Murphy, E. Persson, A. Sivaramakrishnan) and Keck (K. Matthews, B. T. Soifer, and collaborators) telescopes, and to the expert assistance of the staffs of Palomar and Keck Observatories during our observations. I. Smail has kindly provided us with galaxy lists for Abell 2390, as well as assistance in the acquisition of the Abell 851 HST image. This work was supported in part by the NSF PYI award AST-9157412 to S. G. D., and the Greenstein Fellowship to M. A. P.

Table 1. Kormendy Relation Intercepts at $R_e = 1$ kpc

| Cluster | $z$ | $\langle\mu_B\rangle_e$ | $\pm$ | $\langle\mu_R\rangle_e$ | $\pm$ | $\langle\mu_K\rangle_e$ | $\pm$ |
|---|---|---|---|---|---|---|---|
| Coma | 0.024 | 20.19 | 0.12 | 18.84 | 0.06 | 15.63 | 0.11 |
| Abell 2390 | 0.23 | $\cdots$ | $\cdots$ | $\cdots$ | $\cdots$ | 16.01 | 0.19 |
| Abell 851 | 0.407 | 22.95 | 0.16 | 20.30 | 0.12 | 16.36 | 0.09 |

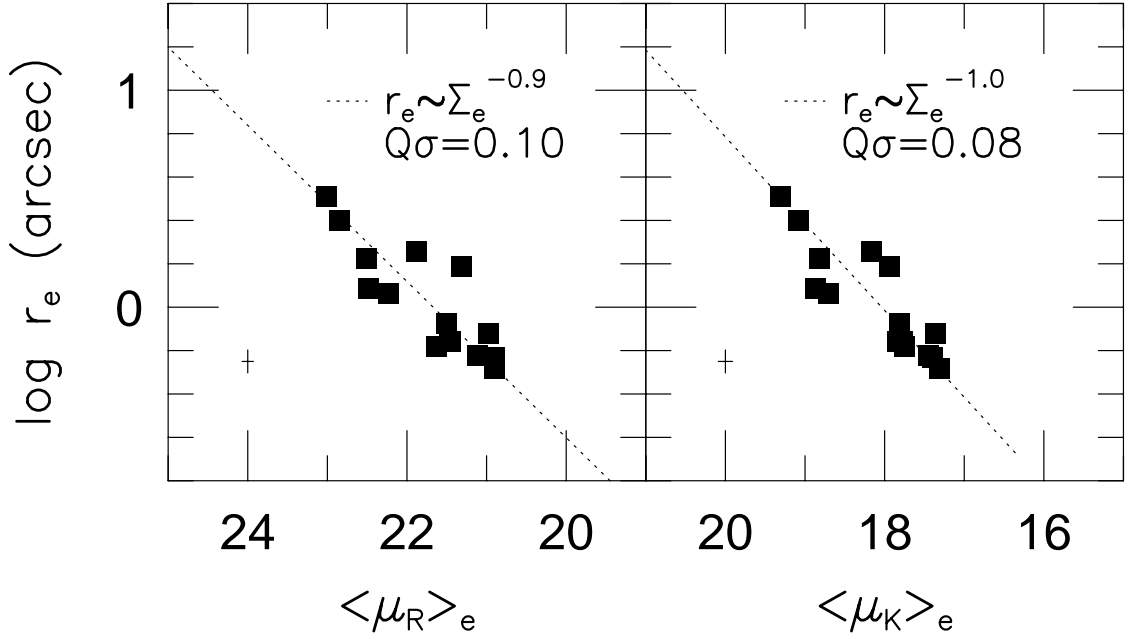

Fig. 1.— Comparison of the Kormendy relations in the (a) $R_C$-band and (b) $K$-band for cluster Abell 851 at $z = 0.407$. The data are well-fit by the Kormendy relation slopes as shown in the figure, with quartile-estimated rms residuals in $\log r_e$ from each relation given as $Q\sigma$.

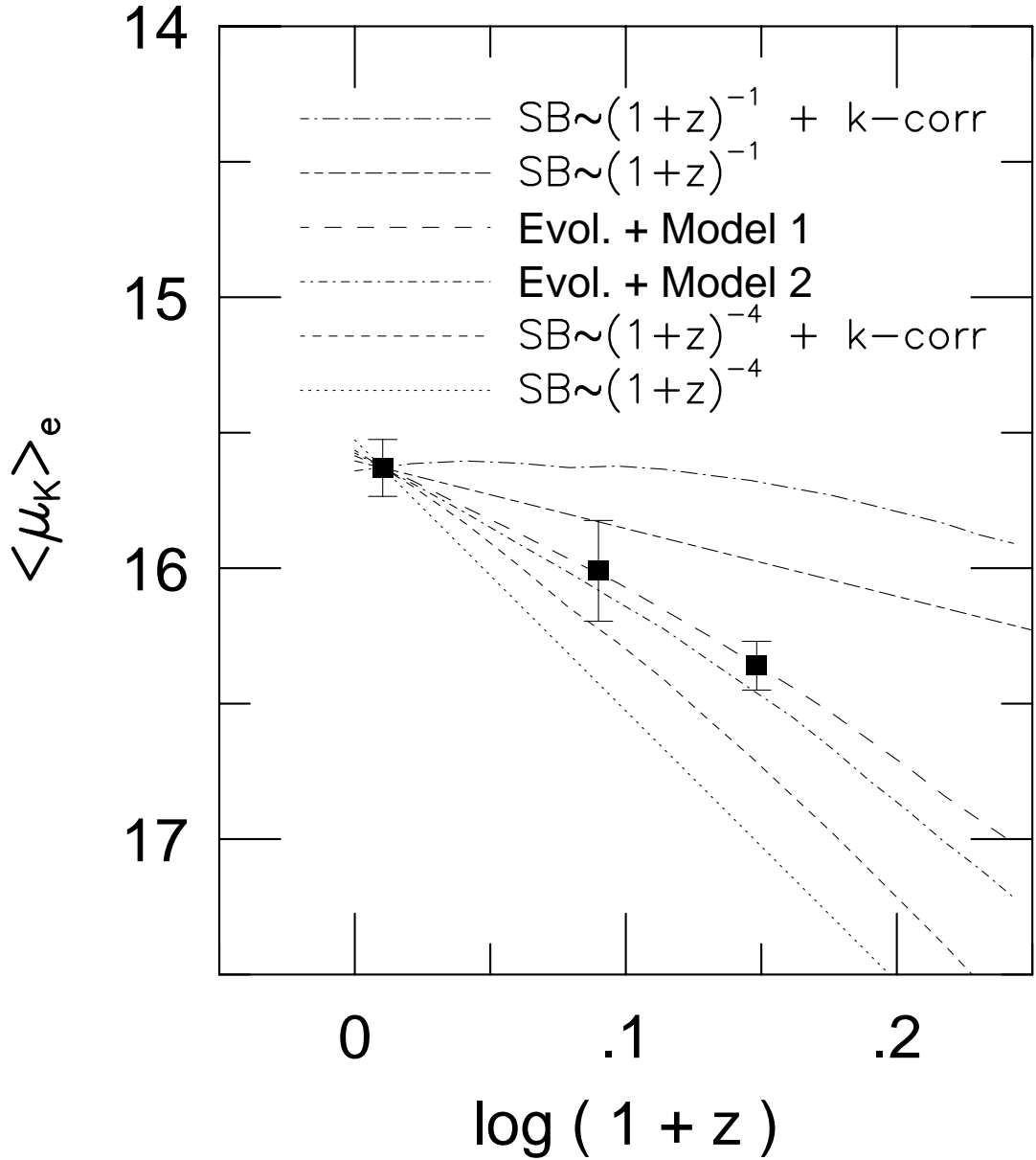

Fig. 2.— The Tolman SB test in the $K$-band using the intercepts on the $\langle\mu\rangle_e$ axis where $R_e = 1$ kpc ($H_0 = 75$ km s$^{-1}$ Mpc$^{-1}$, $\Omega_0 = 0.2$, and $\Lambda_0 = 0$). The data are uncorrected for SB dimming, nor has a $k$-correction been applied. Plotted for comparison are the expansion $(1+z)^{-4}$ and non-expansion $(1+z)^{-1}$ models, both with and without a $k$-correction. Models 1 and 2 (from Buzzoni 1995) both include SB dimming, a $k$-correction, and passive evolution (Evol.) using cosmological parameters ($H_0$, $\Omega_0$, $z_f$) of (50, 0.5, $\infty$) and (100, 0, $\infty$), respectively. The data are consistent with luminosity evolution in the cluster elliptical galaxy population as predicted by the simple stellar population models.

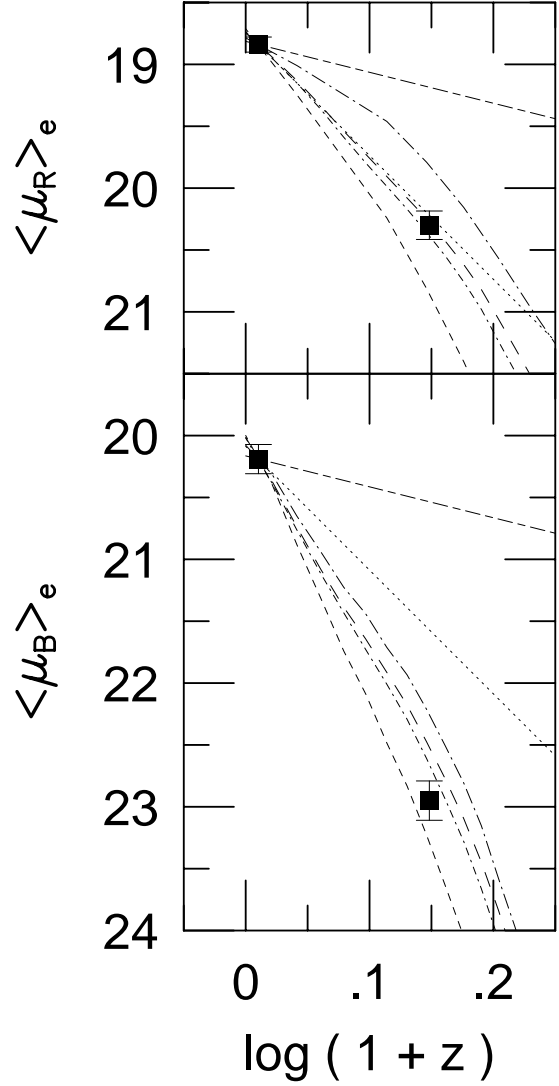

Fig. 3.— The Tolman SB test in the $B$- and $R_C$-bands. The models are as in Figure 2, except that the $R_C$-band $k$-correction is taken from Fukugita, Shimasaku, & Ichikawa (1995). The $R_C$-band signal is consistent with passive evolution of the elliptical galaxy population. The $B$-band signal shows a small deviation from the passive evolution model prediction, which could be due to an imperfect match between the observed and the model bandpasses (either CCD, filter, or both), or to problems in the modelling of the elliptical galaxy spectral energy distribution around 3200Å.